\documentstyle[preprint,aps,harvard]{revtex}
\citationstyle{dcu}
\begin{document}
\title{Classical Monopoles: Newton, NUT-space, gravomagnetic lensing
and atomic spectra} 
\author{D Lynden-Bell$^{1,2}$ \& M
Nouri-Zonoz$^1$} 
\address{\it $^1$Institute of Astronomy, The
Observatories, Cambridge. CB3 0HA\\$^2$Current address Physics
Department, The Queen's University,\\Belfast. BT7 1NN} 
\maketitle
\begin{abstract}
Stimulated by a scholium in Newton's Principia we find some
beautiful results in classical mechanics which can be interpreted in
terms of the orbits in the field of a mass endowed with a
gravomagnetic monopole.  All the orbits lie on cones!  When the cones
are slit open and flattened the orbits are exactly the ellipses and
hyperbolae that one would have obtained without the gravomagnetic
monopole.  

The beauty and simplicity of these results has led us to
explore the similar problems in Atomic Physics when the nuclei have an
added Dirac magnetic monopole.  These problems have been explored by
others and we sketch the derivations and give details of the predicted
spectrum of monopolar hydrogen.  

Finally we return to gravomagnetic monopoles in general relativity.
We explain why NUT space has a non-spherical metric although NUT space
itself is the spherical space-time of a mass with a gravomagnetic
monopole.  We demonstrate that all geodesics in NUT space lie on cones
and use this result to study the gravitational lensing by bodies with
gravomagnetic monopoles.

We remark that just as electromagnetism would have to be extended
beyond Maxwell's equations to allow for magnetic monopoles and their
currents so general relativity would have to be extended to allow torsion for
general distributions of gravomagnetic monopoles and their currents.
Of course if monopoles were never discovered then it would be a
triumph for both Maxwellian Electromagnetism and General Relativity as
they stand!
\end{abstract}
\newpage
\tableofcontents
\newpage
\section{Introduction}  
\label{I}
One of us was asked to review
Chandrasekhar's (1995) book on Newton's Principia (1687) for Notes and
Records of the Royal Society (Lynden-Bell 1996).  This led to reading
passages of Cajori's translation of Principia.  In his first
proposition Newton shows that motion under the influence of a central
force will be in a plane and that equal areas will be swept by the
radius vector in equal times.  In his second proposition he shows that
if a radius from a point $S$ to a body sweeps out equal areas in equal
times then the force is central.  There follows this scholium: ``A
body may be urged by a centripetal force compounded of several forces;
in which case the meaning of the proposition is that the force which
results out of all tends to the point $S$.  But if any force acts
continually in the direction of lines perpendicular to the {\it
described surface}, this force will make the body to deviate from the
plane of its motion; but it will neither augment nor diminish the area
of the {\it described surface} and is therefore to be neglected in the
composition of forces.''

What does this mean?

The words {\it described surface} have been translated from a Latin
word that carries the extra connotation of a surface described by its
edge.  We shall take this to be the surface swept out by the radius
vector to the body that is now describing the non-coplanar path.  A
force normal to this surface at the body must be perpendicular to
${\bf r}$ and ${\bf v}$ which are both within the surface, so Newton
is considering extra forces of the form $Nm_0 {\bf r} \times {\bf v}$
where $N$ may depend on ${\bf r},{\bf v}, t$ etc.  We write the
equation of motion $$m_0 d^2 {\bf r} / dt^2 = -V'(r) {\hat {\bf r}} +
N \,{\bf L}, \eqno (1.1)$$ where $V(r)$ is the potential for the
central force, ${\hat {\bf r}}$ is the unit radial vector, and
$${\bf L} = m_0 {\bf r} \times {\bf v}\,. \eqno (1.2)$$
Taking the cross product ${\hat {\bf r}} \times (1.1)$ we have
$$d {\bf L}/dt = N {\bf r} \times {\bf L} \eqno (1.3)$$
from which it follows either geometrically \`a la Newton or by dotting
with ${\bf L}$ that
$$|{\bf L}| = \ {\rm constant}. \eqno (1.4)$$
Now if $\varphi$ is the angle {\it measured within the described
surface} between a fixed half-line ending at $S$ and the radius
vector,
$${\textstyle{1\over2}} r^2 {\dot \varphi} = {\textstyle{1\over2}}
|{\bf L}|/m_0 \eqno (1.5)$$
so equal areas are swept out in equal times just as Newton says.  To
see this angle more precisely it is perhaps worthwhile to work in axes
which are continually tilting to keep up with the plane of the motion.
In any axes rotating with angular velocity ${\bf \Omega} (t)$, the
apparent acceleration $\ddot {\bf r}$ is related to the absolute
acceleration $d^2 {\bf r}/dt^2$ by
$$d^2 {\bf r}/dt^2 = {\ddot {\bf r}} + 2{\bf \Omega} \times {\dot {\bf
r}} + {\dot {\bf \Omega}} \times {\bf r} + {\bf \Omega} \times ({\bf
\Omega} \times {\bf r})\, . \eqno (1.6)$$
We shall apply this formula to axes which are always tilting about
${\hat {\bf r}}$ such that in these axes the motion appears as planar.
Thus putting ${\bf \Omega} = \Omega {\hat {\bf r}}$ in Eq. (1.6)
$$d^2 {\bf r}/dt^2 = {\ddot {\bf r}} + \Omega r^{-1} {\bf
L}/m_0\,. \eqno (1.7)$$
Inserting this into Eq. (1.1) and choosing
$$\Omega = r N, \eqno (1.8)$$
we recover in these axes the equation we would have had in inertial
axes had Newton's extra force $\propto N$ been absent i.e.,
$$m_0 {\ddot {\bf r}} = -V' {\hat {\bf r}}\,. \eqno (1.9)$$
Thus relative to {\it these moving axes} ${\bf r} \times m_0 {\dot
{\bf r}} = {\bf L}$ is constant not only in magnitude but also in
direction and
$$|{\bf r} \times {\dot {\bf r}}| = r^2 {\dot \varphi} = L/m_0 \eqno (1.10)$$
where $\varphi$ is the angle at $S$ between some line fixed in the moving axes
and the current radial line (this is of course equal to the earlier
angle since this moving plane is `rolling' on the {\it described surface}
about the common radius vector).

We now return to the inertial axes in which the direction of ${\bf L}$
varies in accord with (1.3).  Dotting Eq. (1.1) with ${\bf v} =
d{\bf r}/dt$ the $N$ term goes out so the energy equation is left
unchanged and we have, remembering that $L^2$ is constant,
$${m_0 \over 2} {\bf v}^2 + V = {m_0 \over 2} \left [{\dot r}^2 +
\left ({L \over m_0} \right )^2 r^{-2} \right ] + V = E\, . \eqno (1.11)$$
Here ${\dot r}$ is the same in fixed or rotating axes since this $r$
is scalar.  Equation (1.11) demonstrates that the radial motion $r
(t)$ is precisely that which would have occurred had $N$ been zero.
Furthermore (1.9) and (1.10) demonstrate that within the tilting axes,
or [using (1.5)] within the described surface, the solution $r
(\varphi)$ is precisely the same function that we would have found for
the truly planar motion that occurs with $N$ absent.  Although this
extension of Newton's theorem is not in Principia it would surprise us
if Newton had not seen and understood it.  There is interesting
historical research to be done here on Newton's surviving manuscripts.
We know from Whiteside that this scholium was not in the first draft
of Newton's De Motu Corporum written in Autumn 1684 but appears in its
revision which is probably dated to the Spring of 1685.

Although (1.5) and (1.11) are sufficient for the solution of the
motion within the described surface, we need to find that surface by
solving (1.3) for a complete description of the motion.  This is not
particularly simple and to do it we need to prescribe how $N$ depends
on ${\bf r}, {\bf v}, t$ etc.  However $d{\bf L}/dt$ and $d{\hat {\bf
r}}/dt$ are always parallel since both are perpendicular to ${\bf r}$
and ${\bf L}$.  This led us to consider under what circumstances they
might be proportional.  In particular
$$d{\hat {\bf r}}/dt = -{\hat {\bf r}} \times ({\hat {\bf r}} \times
{\bf v})/r = -{\bf r} \times {\bf L}/(r^3 m_0) \eqno (1.12)$$
so in full generality we have from (1.3)
$$d{\bf L}/dt = -(m_0 N r^3) d{\hat {\bf r}}/dt\, .  \eqno (1.13)$$
This demonstrates that when $m_0 N r^3 = Q_* = \ {\rm constant}$ we
have a beautifully simple solution to (1.13) to wit.
$${\bf L} + Q_* {\hat {\bf r}} = {\bf j} = {\rm const} \eqno (1.14)$$
here ${\bf j}$ is the vector constant of integration; notice that
$Q_*$ has the same dimensions as $L$.  Since ${\bf L}$ and ${\hat {\bf
r}}$ are perpendicular we deduce, dotting with ${\hat {\bf r}}$
$${\bf j} \cdot {\hat {\bf r}} = Q_* \eqno (1.15)$$
which shows that the angle between ${\bf j}$ and ${\hat {\bf r}}$ is
constant so ${\hat {\bf r}}$ moves on a cone whose axis is along ${\bf
j}$.  Similarly dotting (1.14) with ${\bf L}$ we find $L^2 = {\bf j}
\cdot {\bf L}$ so likewise ${\bf L}$ moves on another cone with ${\bf
j}$ as its axis.  If this cone has semi-angle $\chi$ then $L/|{\bf j}|
= \cos \chi$, but ${\hat {\bf r}}$ and ${\bf L}$ are orthogonal and by
(1.14) they are coplanar with ${\bf j}$ so we may choose cf. (1.15)
$$Q_*/|{\bf j}| = \sin \chi \eqno (1.16)$$
so the angle between ${\hat {\bf r}}$ and ${\bf j}$ is $\pi /2 - \chi$
as shown in Fig. 1.  Notice from (1.16) that the angle of the cone
is determined completely from $|{\bf L}|$ and the force constant
$Q_*$.  Orbits with larger $|{\bf j}|$ have smaller $\chi$ so the
angular momentum then moves around a narrow cone and ${\hat {\bf r}}$
then moves around a very open one.  For $|{\bf j}| \gg Q_*$ it is
almost planar.  Fig. 1 illustrates two circular orbits moving in
opposite senses about the same axis.  Notice that the one moving
right-handedly about the upward pointing axis is displaced above the
center sitting like a halo about it while that moving left handedly is
displaced below the center like an Elizabethan ruff below the head.
One might have supposed that for $j \gg Q_*$ these two circular orbits
would approach the central plane but although the cone becomes much
flatter and more open the displacement between the direct and
retrograde orbits actually increases.  For circular orbits at distance
$a$ from $S$ we have, for a Newtonian potential, $L^2 = GMa m_0^2$ and
the displacement is 
$$2{\hat {\bf j}} \cdot {\bf r} = a/ \sqrt{GMa
m_0^2 Q_*^{-2} +1} \rightarrow m_0^{-1} Q_* \sqrt{a/(GM)}.$$
We have been led to the case $m_0 N r^3 = Q_* = \ {\rm constant}$ for
reasons of mathematical simplicity but this case is more than a
mathematical curiosity because:

1. Of all the forces of Newton's $N$ type [see Eq. (1.1)] only those
of the form $-{\bf v} \times {\hat {\bf r}} r^{-2} Q_*(\theta, \phi)$
derive from a Lagrangian.  For a monopole $Q_*$ is constant.

2. We may rewrite this force in the form
$$N {\bf L} = -Q_* {\bf v} \times {\bf r}/r^3 = {m_0 \over c} {\bf v}
\times {\bf B}_g, \eqno (1.17)$$

where $${\bf B}_g = - Q {\hat {\bf r}}/r^2\;\ \ ;\, \ \ Q = Q_*
c/m_0. \eqno (1.18)$$
We have introduced the velocity of light $c$ to make the analogy with
magnetic forces even more obvious.  ${\bf B}_g$ is clearly the field
of a magnetic monopole of strength $Q$ but since this sort of
magnetism acts not on moving charges but rather on moving masses; it
is a gravomagnetic field.  Such fields are well known in general
relativity see Landau \& Lifshitz Theory of Fields (1966) \S89 problem
1.  They are position dependent Coriolis forces associated with what
relativists less helpfully call the dragging of inertial frames.  The
field ${\bf B}_g$ as we have defined it has the same dimensions as
${\bf g}$ the acceleration due to gravity and $Q/G$ has the dimensions
of mass.  In electricity, like charges repel while in gravity, like
masses attract.  It is the same with like magnetic monopoles, they
repel while the gravomagnetic monopoles of like sign attract one
another, hence the negative sign in Eq. (1.18) is best left there
rather than combined into a new definition of the pole strength $Q$.
We may find the Lagrangian corresponding to the force (1.17) by
analogy with the electrodynamic case.  There we add a term $q {\bf
v} \cdot {\bf A}/c$ where $q$ is the charge and ${\bf A}$ is the
vector potential.  For any poloidal axi-symmetric magnetic field one
may choose ${\bf A}$ to be of the form $A {\bf \nabla} \phi$ where
$\phi$ is the azimuth around the axis.  We require
$$-Q {\hat {\bf r}}/r^2 = {\bf B}_g = \nabla \times (A {\bf \nabla}
\phi) = {\bf \nabla}A \times {\bf \nabla} \phi \eqno (1.19)$$
from which one readily finds $A = Q (1+\cos \theta)$ gives the right
${\bf B}_g$.  Thus a Lagrangian for Eq. (1.1) is
$${\cal L} = {\textstyle {1\over 2}} m_0v^2 - m_0 V(r) + Q_* (1 + \cos
\theta) {\bf v} \cdot {\bf \nabla} \phi. \eqno (1.20)$$
Although the dynamical system is spherically symmetrical the
Lagrangian is not and can not be made so.  The only spherically
symmetrical vector fields are $f(r){\bf r}$.  If ${\bf A}$ were of
this form its curl would be zero and therefore
could not be the field of a monopole.  Of course we can choose any
axis we like and measure $\theta$ and $\phi$ appropriately from it.
The ${\bf A}$ field will then be quite different but it will give the
same ${\bf B}_g$ field by construction.  Thus the difference between
any two such ${\bf A}$ fields will have zero curl showing that ${\bf
A}' = {\bf A} + {\bf \nabla} \chi$ i.e., a gauge transformation.  The
Lagrangian (1.20) is neither spherically symmetrical nor gauge
invariant but it is a member of a whole class of equivalent
Lagrangians with different axes which are related by gauge
transformations.  Whereas none of these is individually spherical the
class of all of them is spherically symmetric.  The moral is that it
can be restrictive to impose symmetry on a single member of the class
if the member is not gauge invariant.

So far everything holds for any spherical potential $V(r)$.  We could
for example choose it to be Henon's (1959) isochrone potential
$2aV_0/(a + \sqrt{r^2+a^2})$ or its better known limits the simple
harmonic oscillator $a \gg r$ or the Newtonian potential $a \ll r$.
For all isochrones the orbits can be solved using only trigonometric
functions (see e.g., Lynden-Bell 1963, Evans {\it et al}. 1990).  Here we
shall stick to the Newtonian potential $V/m_0 = -GM/r$.  We have
already shown that the motion lies on a cone whose semi-angle is given
by $\cos^{-1} (Q_*/|{\bf j}|)$; furthermore if we slit that cone along
$\varphi = 0$ and flatten it, the orbit will be exactly what it would
have been in the absence of $N$ i.e., a conic section.  Of course
when we slit and flatten the orbit's cone a gap appears whose angle is
$\gamma = 2 \pi \left [1-L/ \sqrt{L^2 + Q^2_*}\, \right ]$, see Fig.
2.  An ellipse with focus at $S$ and apocentre at $\varphi = 0$ would
get back to apocentre at $\varphi = 2 \pi$ but unfortunately the gap
intervenes.  On the cone we identify $\varphi = 0$ not with $\varphi =
2 \pi$ but rather with $\varphi = 2 \pi - \gamma$.  Thus
on the cone the ellipse will precess forwards by an angle $\gamma$ in
each radial period, Fig. 3.  This angle $\gamma$ is an angle like
$\varphi$ measured at $S$ within the cone's surface.  It is perhaps
more natural to measure angles $\eta$ around the axis of the cone;
these angles are related through ${\dot \eta} = {\dot \varphi}/\cos
\chi = L/(m_0r^2 \cos \chi)$ so $\eta = \varphi \sec \chi = \varphi
|{\bf j}|/L$.

In these terms the precession per radial period is
$$\Delta \eta = 2 \pi (|{\bf j}|/L-1). \eqno (1.21)$$ 
Newton in his proposition on revolving orbits showed that the addition
of an inverse cube force led to an orbit of exactly the same shape but
traced relative to axes that rotate at a rate proportional to $\dot \phi$
in the original orbit.  It is natural to ask whether such an
additional force can stop the precession around the cone of an orbit
in the monopolar problem and so yield an orbit that closes on itself
in fixed axes.  Wonderfully a simple change in $V(r)$ does this not
just for one orbit but for all orbits at once.  We thus obtain a new
superintegrable system in which all bound orbits close.  By analogy
with Hamilton's derivation of his eccentricity vector (Hamilton 1847)
we take the cross product of the equation of motion (1.1) with ${\bf
j} = {\bf L} + Q_* {\hat {\bf r}}$.  On the right hand side two terms
are zero and the remaining two are multiples of $d{\hat {\bf r}}/dt$
cf. (1.12) so we find
$$m_0 {\bf j} \times d^2 {\bf r}/dt^2 = -(m_0 r^2 V' + Q_*^2 r^{-1})
d{\hat {\bf r}}/dt. \eqno (1.22)$$
This will integrate vectorially if the bracket is constant.  Calling
it $GM m_0^2$ we find the potential must be of the form
$$V/m_0 = -GM r^{-1} + {1 \over 2} {Q^2_* \over m^2_0} r^{-2}. \eqno (1.23)$$
Evidently the required inverse cube repulsive force is proportional to
the square of the monopole moment $Q$.  Integrating (1.22) we have
$$d{\bf r} /dt \times {\bf j} = GM m_0 ({\hat {\bf r}} + {\bf e})
\eqno (1.24)$$
where ${\bf e}$ is the vector constant of integration.  Dotting (1.24)
with ${\hat {\bf r}}$ we have
$$\ell_*/r = ({\bf l} + {\bf e} \cdot {\hat {\bf r}}) \eqno (1.25)$$
where $\ell_* = {\bf L} \cdot {\bf j}/(GM m_0^2) = \ {\rm const}$.
Equation (1.25) is the equation of a conic section of eccentricity
${\bf e}$ which defines the direction to pericentre.  But we have not
yet proved that the orbit lies in a plane so (1.25) actually defines
a prolate spheroid, paraboloid or hyperboloid.  Nick Manton, by analogy
with his work on monopoles in Euclidean Taub Space (Gibbons \& Manton
1986), showed us that the motion is in fact planar; for using (1.15)
Eq. (1.25) becomes on multiplication by $Q_*r$
$$Q_* \ell_* = ({\bf j} + Q_* {\bf e}) \cdot {\bf r}$$
which demonstrates that the orbit lies on a plane whose normal is
${\bf j} + Q_* {\bf e}$.  As ${\bf r}$ also lies on a cone this
provides another proof that the motion lies along a conic section.

Notice that the vector integral ${\bf e}$ in (1.24) together with the
integral ${\bf j}$ appears to provide six integrals of the motion.
However they are not all independent because $- {\bf e}\cdot {\bf j} =
{\hat {\bf r}}\cdot {\bf j} = Q_*$.  So they provide 5 independent
integrals.  Thus we have a new superintegrable dynamical system in
which the bound orbits exactly close (cf. Evans 1990, 91).

It was the beauty and simplicity of these results for monopoles in
classical mechanics that led us to believe that a similar simplicity
might well be discernible both in quantum mechanics and in general
relativity.  We were not disappointed, both had already attracted
attention.  Hautot 1972 discusses the separation of variables in $r,
\theta, \phi$ coordinates.  The vector integral ${\bf j}$ is
preferable because the generality of motion on cones is then seen.
For motion in special relativity ${\bf j}$ is still conserved provided
${\bf L}$ is interpreted as $m_0{\bf r} \times d{\bf r}/d\tau =
m_0{\bf r} \times {\bf v}/ \sqrt{1-v^2/c^2}$.  Goddard and Olive
(1978) in their excellent review of monopoles in gauge field theories
quote Poincar\'e (1895) for this integral in the classical case of a
pure electromagnetic monopole.
\section {Dirac's monopole and the spectra of monopolar atoms}
\subsection {Gauge transformations, Schr\"odinger's equation \& Dirac's 
quantised monopole}
The Lagrangian for a particle of mass $m_0$ and charge $-e$ in an
electromagnetic field is
$${\cal L} = {\textstyle {1 \over 2}} m_0 {\dot {\bf r}}^2 - e {\dot
{\bf r}} \cdot {\bf A}/c + e \Phi (r)$$
where $\Phi$ is the electrostatic potential and ${\bf B} = \ {\rm
Curl} \ {\bf A}$ is the magnetic field.  The momentum conjugate to
{\bf r} is
$${\bf p} = \partial {\cal L}/ \partial {\dot r} = m_0 {\dot {\bf r}}
- e {\bf A}/c$$
which is not a gauge invariant quantity.  However the particle's
momentum $m_0 {\dot {\bf r}} = {\bf p} + e {\bf A}/c$ is gauge
invariant and therefore has greater physical significance.  The
Hamiltonian is given by
$$H = {\bf p}\cdot {\bf v} - {\cal L} = (2m_o)^{-1} ({\bf p} + e {\bf
A}/c)^2 - e \Phi . \eqno (2.1)$$
For Schr\"odinger's equation we replace ${\bf p}$ by $-i \hbar {\bf
\nabla}$ and solve $H\psi = E\psi$ for the wave function of a steady
state.  A given magnetic field ${\bf B}$ can be described by many
different vector potentials ${\bf A}$ related by gauge transformations
${\bf A}' = {\bf A} + {\bf \nabla} \chi$.  Each will give us a
different Hamiltonian.  Let us first see how the different wave
functions corresponding to these are related.  Define a new function
$\psi '$ such that
$$\psi = \, {\rm exp} [ie \chi/(\hbar c)] \psi '  \eqno (2.2)$$
then $(-i\hbar {\bf \nabla}+ e {\bf A}/c)\psi = \, {\rm exp}[ie
\chi/(\hbar c)](-i\hbar \nabla + e {\bf A}/c+e{\bf \nabla} \chi /c)
\psi '$ and the combination ${\bf A}' = {\bf A} + \nabla \chi$ has
appeared.  Applying the above operator twice we see that
$$H \psi = \, {\rm exp} [ie \chi/(\hbar c)] H' \psi'$$
where $H'$ is $H$ with ${\bf A}$ replaced by ${\bf A}'$.  It follows
that Schr\"odinger's equation $H \psi = E \psi$ implies $H' \psi' = E
\psi'$ so under gauge transformation $\psi$ transforms to $\psi'$
given by (2.2).

This is a perfectly good wave function whenever $\chi$ is single
valued but following Aharonov and B\"ohm (1959) we now consider the wave
function of a particle outside a small impenetrable cylinder $R=a$.
If we take ${\bf \nabla} \chi = {\bf \nabla}(F\phi /2 \pi), R\geq a$
where $\phi$ is the azimuth this corresponds to the same magnetic
field outside the cylinder but a different magnetic flux within it
because
$$\int{\bf B}' \cdot d {\bf S} = \oint {\bf A}' \cdot d
\mbox{\boldmath $\ell$} = \oint ({\bf A}+\nabla \chi) \cdot d
\mbox{\boldmath $\ell$} = \int {\bf B} \cdot d{\bf S}+F$$ which
identifies the constant $F$ as the extra flux threading the cylinder.
If we adopt our transformation of wave function for a gauge
transformation we get the phase factor
$${\rm exp} [-ie F\phi/(hc)] \eqno (2.3)$$
which is only single valued when the flux takes the special values
$$F=N(hc/e) \eqno (2.4)$$ where $N$ is an integer (positive, negative
or zero).  (This $N$ is not the force coefficient of Section I).  Thus
while we get the correct wave function for those particular values of
$F$, we need to solve the problem anew with the correct boundary
condition that $\psi'$ must be periodic in $\phi$ whenever $F$ is not
an integer multiple of the flux quantum $hc/e$.  Indeed when it is
not, there is interference between the two parts of a beam of
electrons that pass on either side of such a cylinder just because
their phases differ by $e \oint \Delta {\bf A} \cdot d \mbox{\boldmath
$\ell$}/(hc) = eF/(hc)$.  It was just this phase shift that was
observed in the experiments demonstrating the Aharanov B\"ohm effect
of the magnetic flux even when the electron beams were untouched by
the magnetic field.  There is an intimate connection of this result
with Dirac's (1931) earlier quantum of magnetic monopole from which
one flux unit (2.4) emanates..  This comes about because in the
presence of a monopole $\oint {\bf A} \cdot d \mbox{\boldmath $\ell$}$
is itself multivalued.

Consider the integral $\oint {\bf A} \cdot d \mbox{\boldmath $\ell$}$
around a small loop; this is clearly the flux of ${\bf B}$ through the
loop but such a flux is ambiguous in the presence of a monopole since
it depends on whether the surface spanning the loop is chosen to pass
above or below the monopole i.e., $S_1$ or $S_2$ in Fig. 4.  The
difference between these two estimates is just $\int_{S_1 - S_2}{\bf
B} \cdot d{\bf S} = 4 \pi Q$ by Gauss's theorem.  Inserting this
$\Delta \int {\bf A} \cdot d\mbox{\boldmath $\ell$}$ in place of $F$
in (2.3) we see that the wave function will only have an unambiguous
phase provided
$$4 \pi Q=N(hc/e) \eqno (2.5)$$
i.e., provided that the monopole strength is quantized in Dirac units
of ${1 \over 2} \hbar c/e \approx {137e \over 2}$.

The quantum of magnetic flux (2.4) is inversely proportional to the
charge.  Quanta of half this size are observed in the Josephson effect
in superconductivity where the effect is due to paired electrons.
There is some evidence for the larger unit (2.4) in ordinary conductors
at low temperatures (Umbach {\it et al}. 1986).

Returning to Schr\"odinger's equation (2.1) and using the vector potential
[cf. under (1.19)]
$${\bf A} = -Q(1+\cos \theta)\nabla \phi \eqno (2.6)$$ we have the
correct magnetic field for a monopole of strength $Q$ but we notice
that ${\bf A}$ is singular along the line $\theta = 0$ although it is
regular along $\theta = \pi$.  Near the singular line ${\bf A}
\rightarrow -2Q \nabla \phi$ which is the vector potential of a tube
carrying a flux $4 \pi Q$ downwards, thus formula (2.6) represents the
vector potential of a magnetic monopole fed its flux by the singular
half line $\theta = 0$.  This half line gives an unobservable
Aharanov-B\"ohm effect provided $4\pi Q = Nhc/e$ that is provided the
monopole is a multiple of the Dirac (1931, 48) unit.  Extra interest
in his monopole comes from his argument that it can also be read as a
reason for charge quantization, because, if $Q$ is the least monopole,
then $e$ must be a multiple of $\hbar c/Q$; thus in his picture,
charge quantization and monopole quantization spring from the same
source.  It is of interest that ${\bf A}$ in (2.6) is single valued.
It has avoided the multi-valuedness alluded to above by having the
singular string at $\theta = 0$ down to the monopole.  This plays the
role of the cut in multivalued functions in the complex plane Wat \&
Yang (1976).  An interesting historical remark is that Schr\"odinger
in 1922 saw that quantum conditions in the old quantum theory led to
$\Gamma \equiv {\textstyle {e \over c}}[\oint \Phi dt -{\bf A} \cdot
d{\bf x}] = nh$ while Weyl's gauge theory led him to consider exp
$(-\Gamma/ \gamma)$ with $\gamma$ as yet unspecified.  He realized
that the identification $\gamma = -i\hbar$ would lead naturally to
such quantum numbers and after de Br\"oglie (1925), he built on this
idea to invent his wave mechanics in 1926.  (See Yang 1987).

\subsection {Solution of Schr\"odinger's Equation}

Written in spherical polar coordinates Schr\"odinger's equation is
\begin{eqnarray*}
-{\hbar ^2\over2 m_0 r^2} \left\{ {\partial \over\partial r} \left(r^2
{\partial \psi \over\partial r} \right) + {\partial\over\partial\mu}
\left[(1-\mu^2) {\partial\psi\over\partial\mu} \right] + \right. \\
\mbox{}+ \left. {1\over 1-\mu^2}
\left[{\partial^2\psi\over\partial\phi^2} - iN (\mu+1)
{\partial\psi\over\partial\phi} - {\textstyle {1\over 4}}N^2 (\mu +
1)^2 \psi \right] \right \} - \end{eqnarray*}$$- e\, \Phi\, \psi = E \psi 
\eqno (2.7)$$
here $\mu$ has been written for $\cos \theta$ and
$N$ is the number of Dirac monopoles on the nucleus.  $\phi$ only
occurs as $\partial / \partial\phi$ in the above equation so we may
take one Fourier component with $\psi \propto \ {\rm exp}\, (im \phi)$
and $m$ an integer positive, negative or zero in order that $\psi$ be
single valued.  On multiplication by $-2 m_0 r^2/(\hbar ^2 \psi)$ we
then find the separated equation
$$\displaylines{{1 \over \psi} {\partial \over \partial \mu} \left[(1-
\mu^2) {\partial \psi \over \partial \mu} \right] - {[m-N {\textstyle
{1 \over 2}} (\mu +1)]^2 \over 1-\mu ^2} = -C = \cr \hfill{} = -{1
\over \psi} {\partial \over \partial r} \left(r^2 {\partial \psi \over
\partial r}\right) - {r^2 2m_0 \over
\hbar^2}(E+e\Phi). \hfill{(2.8)}\cr}$$ Writing $\psi=\psi_r(r)
\psi_\mu (\mu)$ the left hand side is a function of $\mu$ alone and
the right hand side is a function of $r$ alone so both must be a
constant which we call $-C$.  The resultant equation for $\psi_\mu$
has regular singular points at $\mu=\pm 1$.  The indicial equations
for the series solutions about $\mu =\pm 1$ have regular solutions
behaving as $(1-\mu)^{{\textstyle {1 \over 2}}|m-N|}$ and
$(1+\mu)^{{\textstyle {1 \over 2}}|m|}$ respectively, so we remove
those factors by writing $$\psi_\mu = (1-\mu)^{{\textstyle {1 \over
2}}|m-N|} (1+\mu)^{{\textstyle {1 \over 2}}|m|} F(\mu). \eqno
(2.9)$$After some algebra the equation for $F$ takes the form
$$\displaylines{(1-\mu^2)F'' + [ (|m| + 1)(1-\mu) -
(|m-N|+1)(1+\mu)]F' +\cr +{\textstyle {1 \over 2}} [2C - m(m-N) -
|m||m-N|-|m-N|-|m|] F = 0.\cr \hfill {(2.10)}\cr}$$

We now write $z = {\textstyle {1\over 2}}(1+\mu)$, so
$z(1-z)=(1-\mu^2)/4$ and $dz={\textstyle {1\over2}} d \mu$ which
reduces the above equation into the standard form for the
hypergeometric equation i.e.,
$$z(1-z)d^2 F/dz^2 + [c-(a+b+1)z] \, dF/dz - abF=0 \eqno (2.11)$$
where $$c = |m| + 1 \eqno (2.12)$$ 
$$a+b = |m| + |m-N| +1 \eqno (2.13)$$
and $-2ab$ is the final square bracket in Eq. (2.10).

The hypergeometric function finite at $\mu = -1, z=0$ diverges like
$(1-\mu)^{c-a-b}$ at $\mu =1$, that is twice as fast as the first
factor in (2.9) converges, so in order to get convergence the
hypergeometric series must terminate.  This occurs only if $a$ or $b$
is a negative integer or zero.  w.l.g. taking $b=-K$ we find that $F$
reduces to a Jacobi polynomial $P^{\alpha \beta}_K (\mu)$ so that
$\psi_\mu$ takes the form
$$\psi_\mu = C_{kmn}(1-\mu)^{1/2 |m-N|}(1+\mu)^{1/2 |m|}
P^{|m-N|,|m|}_K \eqno (2.14)$$
here $\int^{+1}_{-1} \psi^2_\mu d\mu = 1$ and $C_{kmn}$ is the normalization
$$\left[{(2K+|m-N|+|m|+1) K! (K+|m-N|+|m|)! \over
2^{|m-N|+|m|+1} (K+|m-N|)! (K+|m|)!} \right]^{{\textstyle{1\over2}}}.$$
The condition that $b=-K$ gives $$-2ab = 2K(|m| +|m-N| +1+K)$$ and hence
(noticing that $K=0$ leaves (2.14) finite) we have
$$\displaylines{C=K(K+1)+K(|m| +|m-N|) + \cr \hfill{}
+{\textstyle {1\over
2}}\left[m(m-N)+|m||m-N|+|m-N|+|m|\, \right]. \hfill{(2.15)}\cr}$$
If we write $j=K+ {\textstyle {1 \over 2}}(|m| +|m-N|)$, then we
notice that $j$ is a positive half-integer and $j\geq {\textstyle {1
\over 2}}(|m| + |m-N|)$
$$C = j(j+1) - N^2/4. \eqno (2.16)$$
Thus $C$ and $j$ are only integers when $N$ is an {\it even} integer.
When $N$ is odd $C$ and $j$ are an integer $\pm {\textstyle {1 \over
2}}$.  For given $j$ and $N \geq 0, \ \ \ m- {\textstyle {N \over 2}}$
takes the $2 j+1$ values from $-j$ to $+j$ in steps of 1.  For $N=1$
the ground state has $j={\textstyle {1\over 2}}$ and $C= {\textstyle
{1\over 2}}$ rather than the values $0$ familiar from the normal
hydrogen atom.  The $j={\textstyle {1\over 2}}$ states with $m=1$ and
$m=0$ are degenerate, see Fig. 5.

With the value (2.16) for $C$ we now turn to the radial equation for
$\psi_r$, (2.8).  Here the treatment is very close to the classical
case clearly laid out by Pauling and Wilson (1935).  We take $\Phi =
Ze/r, Ze$ being the nuclear charge, and $E$ negative.  We write
$$\alpha^2 = -2m_0 E/\hbar ^2 \eqno (2.17)$$
$$\zeta = m_0 Z e^2 h^{-2} \alpha^{-1} \eqno (2.18)$$
and use a normalized radius $\tilde {r} = 2\alpha r$.  As all the
radii in the rest of this section are so normalized we shall forget
the \~{ \ }and take it as read.  Eq. (2.8) now takes the form
$${1 \over r^2} {d \over dr} \left (r^2 {d\psi_r \over dr}\right ) +
\left (-{1\over 4} - {C \over r^2} + {\zeta \over r} \right ) \psi_r =
0. \eqno (2.19)$$
The asymptotic form of this equation for large $r$ is $\psi_r'' =
\psi_r/4$ so $\psi_r \rightarrow \, {\rm exp} (r/2)$ or ${\rm exp} -
(r/2)$.  Of these only the second is acceptable so we write $$\psi_r =
\, {\rm exp} (-r/2) r^s f(r) \eqno (2.20)$$ where $f(r)$ may be
expanded in series about the origin in the form $$\sum^\infty_{p=0}\,
a_p\, r^p$$ and $s$ is chosen so that $a_0 \not=0$.  The indicial
equation found by substitution of the series (2.20) into (2.19) is
$$[s(s+1) -C]a_0 =0$$
but by hypothesis $a_0 \not=0$ so using (2.16) $s$ is given by 
$$\left (s+{\textstyle {1 \over 2}}\right )^2 = {\textstyle {1 \over
4}}+s \left (s+1 \right)={\textstyle {1 \over 4}} + C = \left
(j+{\textstyle {1 \over 2}}\right )^2 - {\textstyle {1\over 4}} N^2.
\eqno (2.21)$$
The recurrence relation for general $p$ is then
$$p(p+2s+1)a_p = (s+p-\zeta)a_{p-1}$$
and the asymptotic form for large $p$ is $a_p \rightarrow a_{p-1}/p$
which shows that $f \rightarrow e^r$.  In that case $\psi _r$ would
diverge at large $r$.  This is unacceptable so the series must
terminate.  Thus there must be a positive integer $p=n'+1$ such that
$$\zeta = p+s =n'+s+1 \eqno (2.22)$$
with $s$ given by (2.21).  Returning to (2.18) and (2.17) this gives the
eigenvalues for the energy in the form 
$$E = -{m_0 Z^2e^4 \over 2\hbar ^2} {1 \over (n'+s+1)^2} = -{m_0Z^2e^4
\over 2 \hbar ^2} {1 \over (n+ \Delta)^2} \eqno (2.23)$$
where $n=n'+J+1$ where $J$ takes values 0, 1, 2 $\ldots$ replaces the
usual $\mbox{\boldmath $\ell$}$ and
$$J = j - {\textstyle {1 \over 2}}|N| \geq 0$$
$N$ is the number of Dirac monopoles on the nucleus and
$$\Delta = \sqrt{(J+{\textstyle {1\over2}}) (J+{\textstyle {1\over2}}
+ |N|)} - \left(J+{\textstyle {1\over2}} \right) \ \ \ \geq 0.$$
Notice that $\Delta$ depends on $J$ as well as $|N|$ and is only zero
when $N = 0$.
For large $J/|N|$, $$\Delta \rightarrow {\textstyle {1 \over 2}}|N|
\left[ 1- {1\over 4} {|N| \over J+{\textstyle {1 \over 2}}} + {1 \over
8} \left( {|N| \over J+ {\textstyle {1 \over 2}}} \right) ^2 - \cdots
\right],$$
while for the ground state $J = 0$
$$\Delta = {\textstyle {1 \over 2}} \left( \sqrt{2|N|+1} - 1 \right)$$
which becomes ${\textstyle {1 \over 2}} \left( \sqrt{3} -1 \right)$
for $N=1$.  So $\Delta$ is {\it not} small.  For a spinless electron
the degeneracy of a state of given $J$ and $n$ is $2j+1 = 2J+1+|N|$
with $m-{\textstyle {1 \over 2}} |N|$ taking values from $-j$ to $+j$.
Notice that the ground state $J=0,n=1$ is a doublet for $N = 1$ and
has $j$ value ${\textstyle {1\over 2}}$ with $m = 0$ and $m=+1$ states
even {\it before} we have allowed for further degeneracy due to
electron spin.  A single Dirac monopole thus gives some effects
reminiscent of spin ${\textstyle {1\over 2}}$ particles (Goldhaber 1976).

The dependence of $\Delta$ upon $J$ lifts the degeneracy of the
different $J$ states ($\ell$ states) that occurs in normal hydrogen.
The energy levels are near to those for an atom with a true spinning
electron laid out in Figs. 6, 7, 8 and Tables 1 \& 2.  The
degeneracy would return if the extra repulsive potential ${\textstyle
{1\over 2}} Q^2/(m_0r^2c^2)$ were included.  Then the $-{\textstyle {1
\over 4}}N^2$ in (2.21) would be cancelled so $s$ would become equal
to $j$.

\subsection {Selection Rules}

The string to the monopole makes it look non-spherical but this is not
truly the case as putting the string in any other direction can be
achieved by a mere gauge transformation.  Therefore without loss of
generality we may evaluate transition moments by taking the
displacement in the $z$ direction in which case we get
$$\displaylines{ R_{ab} = \int \psi^*_a r \mu \psi_b d^3 r= \cr = 2\pi
\delta_{m_am_b} \int^\infty _0 \psi_{ra} \psi_{rb} r^3 dr
\int^{+1}_{-1} \psi_{\mu a} \mu \psi_{\mu b} d\mu. \cr}$$ The radial
integral is that for normal hydrogen but the scales have changed since
$s$ in (2.22) is no longer $\ell$ but is given instead by (2.21).  We
shall concentrate on the important change in selection rules given by
the final integral.

Whereas for the Legendre Polynomials in normal hydrogen wave functions
we have
$$\mu P_\ell (\mu) = {\ell + 1 \over 2\ell +1} P_{\ell +1} + {\ell
\over 2\ell +1} P_{\ell -1}$$
so that $\int^{+1}_{-1} P_{\ell '} \mu P_\ell d\mu$ is only non-zero
when $\ell' - \ell = \pm 1$, for the Jacobi polynomials in monopolar
hydrogen
$$\mu P_K^{\alpha \beta} = \left (a_1 P_{K+1}^{\alpha \beta} + a_2
P^{\alpha \beta}_K +a_4 P^{\alpha \beta}_{K-1} \right)/a_3$$
where 
\begin{eqnarray} 
a_1 & = & 2(K+1)(K+\alpha + \beta +1)(2K +\alpha + \beta) \nonumber \\ 
a_2 & = & (2K + \alpha + \beta +1)(\alpha^2 - \beta^2) \nonumber \\ 
a_3 & = & (2K +\alpha +\beta)(2K + \alpha + \beta +1)(2K + \alpha + \beta
+2) \nonumber \\ 
a_4 & = & 2(K+\alpha) (K+\beta)(2K+\alpha +\beta +2) \nonumber
\end{eqnarray}
so that $\int ^{+1}_{-1} (1-\mu)^\alpha (1+\mu)^\beta P^{\alpha
\beta}_{K'} \mu P_K^{\alpha \beta} d\mu$ will be non-zero when $K' - K
=\pm 1 \, {\rm or} \, 0$.  [The 0 term is only absent when $a_2 =0$
i.e., $\alpha \equiv |m-N| = \beta \equiv |m|$.  This occurs for $N=0$
always, for $N=2$ when $m=1$, but never for $N=1$.]

Thus there is a significant change in the selection rules for electric
dipole transitions.  Some might imagine that magnetic dipole
transitions should be important but the magnetic monopole is on a
heavy nucleus and barely responds to an oscillating magnetic field so
it is still the electric dipole transitions of the electron that are
important.  $m$ is unchanged for a dipole along the z-axis so $\Delta
K=\pm 1$ or $0$ leads directly to $\Delta j$ and hence \break $\Delta J = \pm
1 \ {\rm or} \ 0$ for such transitions.

Even order of magnitude estimates show that the interaction of the
electron spin's magnetic moment with unit monopole gives not a
delicate fine structure but significant changes in the eigenvalues!
Thus to find the true eigenvalues the Dirac equation is a necessity!
Before treating it we clear up some details.  We took the form (2.2)
for ${\bf A}$ corresponding to a monopole with a string along $\mu
=+1$.  For $|N| \geq 2$ we could have taken two or more inwardly
directed strings of flux.  Are these different string configurations
really different monopoles or do they all give the same eigenvalues?
The effect of such a change is to add a unit flux string along the $z$
axis.  It is simple to show that this is equivalent to adding one to
$m$ everywhere that it occurs.  Provided we do that also to $m$ in the
definition of $j$ under (2.15) the final spectrum remains unchanged.
What does change are the $K$ and $m$ values associated with a given
$j$ value.

A second detail is the value of $m_0$ which for $N=0$ would be the reduced
mass of the electron so for hydrogen it is $m_0 = m_em_p/(m_e+m_p)$.

Particle physicists expect a heavy mass for any monopole so any
monopolar hydrogen will have a nucleus much heavier than the proton
and $m_e$ should be substituted for $m_0$ in predicting spectra.  A third
detail for later reference is the energy spectrum of the relativistic
Klein-Gordon equation.  Here we follow Schiff's
treatment and obtain writing $\alpha_z = Ze^2/(\hbar c)$
$$E = m_0c^2 \left\{ \left[ 1+ {\alpha^2_z \over (n+\Delta_1)^2}
\right] ^{-1/2}-1 \right\} \eqno (2.25)$$
where $$\Delta_1 = \sqrt{(J+{\textstyle {1 \over 2}})(J+ {\textstyle
{1\over 2}}+|N|) - \alpha^2_z} - \left( J+{\textstyle {1 \over 2}}
\right). \eqno (2.26)$$

\subsection {Angular Momentum}

Returning to the classical conserved quantity cf. (1.14) we see the
conserved quantity is not the particle's angular momentum ${\bf L} =
{\bf r} \times m_0 {\bf v}$ but rather that supplemented by $eQc^{-1}
{\hat {\bf r}}$.  The physics behind this supplement lies in the
Poynting vector of the electromagnetic field which carries an angular
momentum

$${}$$
\begin{eqnarray*}{1 \over 4\pi c} \int {\bf r}' \times \left( {\bf E} \times 
{Q \over r'^2} {\hat {\bf r}}' \right) d^3r' = {Q \over 4 \pi c} \int \left(
{\bf E} \cdot {\bf \nabla} \right) {\hat {\bf r}}' d^3r' = \\ \mbox{}
= {-Q \over 4 \pi c} \int {\hat {\bf r}}' {\bf \nabla} \cdot {\bf E}
d^3 r' = + {eQ \over c} {\hat {\bf r}} = + {\textstyle {1\over 2}} N
\hbar {\hat {\bf r}}\end{eqnarray*} where ${\bf \nabla} \cdot {\bf E}
= - e4 \pi \delta^3 ({\bf r}' - {\bf r})$.  The total angular momentum
is thus ${\bf j} = {\bf L} + eQc^{-1} {\hat {\bf r}}$.

As we saw above (2.1) $m_0{\bf v} = {\bf p} + e {\bf A}/c$ in the
presence of a magnetic field so the operator representing ${\bf j}$ is
${\bf r} \times \left( -i \hbar {\bf \nabla} + e{\bf A}/c \right) +
{\textstyle {1\over 2}} N \hbar {\hat {\bf r}}$.  The commutators
$\left[ -i\hbar \partial_j + e A_j/c, - i\hbar \partial_k + eA_k/c
\right] = -i \hbar e\, c^{-1} \varepsilon_{jkl} B_l = -i \hbar
eQc^{-1} \varepsilon_{jkl} x^l/r^3$ and $[-i \hbar \partial_j + eA_j/c,
x^k] = -i \hbar \partial_j^k$ enable one to derive the commutator
$$[j_j,j_k] = i \hbar \varepsilon_{jkl}j_l$$
which demonstrates that ${\bf j}$ obeys the angular-momentum algebra
of the rotation group.  One may also demonstrate that ${\bf j}^2$
commutes with ${\bf j}$ and that $j_{\pm} = j_x \pm ij_y$ are the raising
and the lowering operators for $j_z$.  From which it follows by the
usual argument that the eigenvalues of $j_z$ are $-j \hbar$ to $+j
\hbar$ and that the eigenvalues of ${\bf j}^2$ are $j(j+1) \hbar ^2$.
But $|{\bf j}|^2 = |{\bf L}|^2 + {\textstyle {1 \over 4}} N^2 \hbar
^2$ so the eigenvalues of $|{\bf L}|^2$ are $\left[ j(j+1) -
{\textstyle {1 \over 4}} N^2 \right] \hbar ^2$.  Now looking at our
separation of variables expression (2.8) we see that the LHS is just
$-\hbar ^{-2} |{\bf L}^2|$ by construction and hence $C = \left( j
(j+1) - {\textstyle {1\over 4}}N^2 \right)$ which agrees with (2.16)
and identifies the $j$ defined there with the generalized angular
momentum eigenvalue defined in this section.  Note that for a single
Dirac monopole and a non-spinning electron we showed (2.16) that $j$
took {\it half odd integer values}.

In the next section we look at the Dirac equation for a spinning
electron.  There the correct generalization is ${\bf j} = {\bf L} +
{\textstyle {1 \over 2}} N \hbar {\hat {\bf r}} + {\textstyle {1\over
2}} \hbar \mbox{\boldmath $\sigma$}$.

This new ${\bf j}$ obeys the angular momentum algebra of the rotation
group but now its eigenvalues are $j(j+1) \hbar ^2$ with $j$ taking
integer (or half integer) values $\geq {N+1 \over 2}$ depending on
whether $N$ is odd or even.

\subsection {Dirac Equation}

The Dirac equation may be written in standard notation $$H\psi =
\left[ -c \mbox{\boldmath $\alpha$} \cdot ({\bf p} + e{\bf A}/c) -
\beta m_0c^2 + V \right] \psi = E\psi$$ with the newly defined ${\bf
j}$ we find ${d{\bf j} \over dt} = [{\bf j}, H] = 0$ so that each
component of this generalized ${\bf j}$ commutes with the Hamiltonian
always provided that ${\bf A}$ is a vector potential for the monopole.
Following Schiff's treatment (1955) we define $p_r = r^{-1} ({\bf r}
\cdot {\bf p} - i\hbar)$ and $\alpha r = r^{-1}(\mbox{\boldmath
$\alpha$} \cdot {\bf r})$ and $\hbar {\bf k} = \beta \sigma' \cdot
\left( {\bf r} \times \left( {\bf p} + {e {\bf A} \over c} \right) +
\hbar \right)$.  No ${\bf A}$ term is needed in $p_r$ since ${\bf r}
\cdot {\bf A} = 0$ for our monopole.

The Hamiltonian is now written
$$H = -c \alpha_r p_r -i {\hbar c \over r} \alpha_r \beta k - \beta
m_0c^2 + V$$
and as before $\alpha_r , \beta$ and $p_r$ all commute with ${\bf k}$.
The eigenvalues of ${\bf k}$ follow by squaring the definition
$$\hbar ^2 k^2 = (\mbox{\boldmath $ \sigma$}' \cdot {\bf L})^2 + 2
\hbar \mbox{\boldmath $\sigma$}' \cdot {\bf L} + h^2 = L^2 +
{\textstyle {1 \over 4}} \hbar ^2.$$ In the last section we showed
that $L^2$ has eigenvalues $j (j+1) \hbar ^2 - {N^2 \over 4} \hbar ^2$
where $j$ was an integer ($N$ odd) or half odd integer ($N$ even) so
$k^2$ has eigenvalues $\left( j+{\textstyle {1 \over 2}} \right)^2 -
{\textstyle {1 \over 4}} N^2$.  Save for this change of $k$ the usual
separation of the Dirac equation goes through unscathed and following
Schiff one obtains the energy levels
$$E = m_0c^2 \left\{ \left[ 1+ {\alpha^2_z \over (s + n')^2}
\right]^{-1/2} -1 \right\}$$
where $s = \left( k^2 - \alpha^2_z \right)^{1/2}$ and $\alpha_z =
Ze^2/(\hbar c)$.

$n'$ is the radial quantum number.  Inserting our eigenvalues $k^2 =
\left(j + {\textstyle {1\over 2}} \right)^2 - {\textstyle {1\over 4}}
N^2$ with $j = J + {\textstyle {1 \over 2}} (|N|+1)$ and $J=0,1,2$,
etc. we have
$$E = m_0c^2 \left\{ \left[1 + {\alpha ^2_z \over (n + \Delta)^2}
\right]^{-1/2} -1 \right\}$$
where $n = n'+J+1$ and
$$\Delta = \sqrt{(J+1)(J+1+|N|) - \alpha^2_z} - (J+1).$$ These energy
levels were first derived by Hautot (1972, 73) generalizing
Harish-Chandra's (1968) separation of the Dirac equation\footnote{For
the further generalization to the problem with an additional
B\"ohm-Aharonov string, see Villalba (1994, 95) \& Hoa\"ng {\it et
al}. (1992).}.  For the scattering by monopoles see also Goldhaber
(1965), Kazama \& Yang (1976) and Kazama {\it et al}. (1977).  We have
drawn the bound energy levels that result Figs. 6, 7 and 8 and derived
the wavelengths of the lines of ``Monopolar Hydrogen'' with one or two
Dirac monopoles attached to the nucleus.  Tables 1 \& 2.  Schwinger
(1966) has suggested that the unit monopole should have the strength
of two Dirac monopoles.  With colleagues he has also calculated the
motions of charged monopoles, dyons, under their mutual attraction
(Schwinger {\it et al}. 1976).  While monopoles may seem esoteric it
is worthwhile looking for lines of monopolar hydrogen in the spectra
of exotic astronomical objects.
 
\section {Gravomagnetic Monopoles in General Relativity, NUT Space}

\subsection {NUT space the general spherically symmetric gravity field}

Zelmanov (1956) and Landau \& Lifshitz (1966) in developing their very
physical approach to general relativity consider stationary
space-times and put the metric in the form
$$ds^2 = e^{-2 \nu} (dx^0 - A_\alpha dx^\alpha)^2 - \gamma_{\alpha,
\beta} dx^\alpha dx^\beta \eqno (3.1)$$ where $\nu \geq 0,A_\alpha$
and $\gamma_{\alpha \beta}$ are independent of $x^0 = ct$.  (Our $\nu$
is $-{\textstyle {1 \over 2}} \nu$ of Landau \& Lifshitz).

However this form is not unique since a transformation of time zero
$x'^0 = x^0 + \chi (x^\alpha)$ leads to
$$ds^2 = e^{-2 \nu} (dx'^0 - A'_\alpha dx^\alpha)^2 - \gamma_{\alpha
\beta} dx^\alpha dx^\beta$$
where $A'_\alpha = A_\alpha + \nabla_\alpha \chi$ so under such a
change ${\bf A}$ undergoes a gauge transformation.  Landau \& Lifshitz
also show that $\gamma_{\alpha \beta}$ can be regarded as a metric of
space i.e., the quotient space $V^4/L^1$ (Geroch 1970) -- as opposed
to space-time.  They show that test bodies following geodesics of
space-time depart from the geodesics of space as if acted on by
gravitational forces which in our notation take the form
$${\bf f} = {m_0 \over \sqrt{1-v^2/c^2}} \left[{\bf E}_g + {{\bf v}
\over c} \times e^{- \nu} {\bf B}_g \right] \eqno (3.2)$$
where the gravitational field $${\bf E}_g = c^2 {\bf \nabla}\nu \eqno
(3.3)$$ and $${\bf B}_g = c^2 \ {\rm Curl} \ {\bf A}. \eqno (3.4)$$  
The conserved energy of the particle in motion is
$$\varepsilon = m_0c^2e^{-\nu} \left( 1-v^2/c^2 \right)^{-1/2} \eqno (3.5)$$
$e^{-\nu} <1$ is the redshift factor by which energy is degraded.

Rewriting Landau \& Lifshitz's form of Einstein's equations (\S95
problem) we find
$${\rm div} \ {\bf B}_g = 0 \eqno (3.6)$$
$${\rm Curl} \ {\bf E}_g = 0 \eqno (3.7)$$
$$\displaylines{{\rm div} \ {\bf E}_g = -c^{-2} \left[ 4\pi G {(\rho
c^2+3p) + {v^2 \over c^2} (\rho c^2-p) \over 1-v^2/c^2}\right. - \cr
\hfill{} \left. -{\textstyle {1 \over 2}} e^{-2 \nu} {\bf B}_g^2 -
{\bf E}_g^2 \Biggr] \right. \hfill{(3.8)} \cr}$$ where $\rho$ is the
energy density in the rest frame of the fluid, $3p$ is the trace of
its pressure tensor and $v$ its velocity defined locally by local time
synchronized along the fluid's motion.  For non-relativistic
velocities this equation reduces to Poisson's equation with the
primary term on the right being $4 \pi G\rho$.  The remaining term has
the form of a negative energy density contributed by the gravity
fields.  The next equation takes the form
$${\rm Curl} \ (e^{-\nu} {\bf B}_g) = -c^{-3} [16 \pi G {\bf j}_g - 2c
{\bf E}_g \times e^{- \nu} {\bf B}_g]. \eqno (3.9)$$
Notice that $e^{- \nu} {\bf B}_g$ occurs also in that combination in the
expression for the force.  It is attractive to regard the final term
as an energy current corresponding to a Poynting vector flux of
gravitational field energy.  ${\bf j}_g$ the matter energy current is
given by
$${\bf j}_g = {\rho c^2 + p \over 1 - v^2/c^2} \ {\bf v}.$$
The final Landau \& Lifshitz equation for the 3 stress tensor is
$$\displaylines{P^{\alpha \beta} - E_g^{\alpha; \beta} = \left(
T^{\alpha \beta} + {\textstyle {1 \over 2}} {\dot T} \gamma^{\alpha
\beta} \right) + \cr \hfill{}+e^{-2 \nu} (B^\alpha _g B^\beta _g -
B^2_g \gamma ^{\alpha \beta}) + E^\alpha _g E^\beta _g,
\hfill{(3.10)}\cr}$$ $P^{\alpha \beta}$ is the 3 dimensional Ricci
Tensor constructed from the metric $\gamma^{\alpha \beta}$.  Those
familiar with the Maxwell stresses of magnetic and electric fields in
say magnetohydrodynamics will find some interest in the field terms on
the right.  The matter terms may be rewritten as physical quantities
for an isotropic fluid in motion
$$T^{\alpha \beta} + {\textstyle {1 \over 2}} T \gamma^{\alpha \beta}
= {8 \pi G \over c^4} \left[ {(p + \rho c^2) v^\alpha v^\beta \over c^2-v^2}
+ {\textstyle {1 \over 2}} (\rho c^2 - p) \gamma^{\alpha \beta}
\right].$$
It should be stressed that all these equations hold good even when
space-time is strongly curved.  Unlike some treatments they are not
restricted to nearly flat-space but it is assumed that the space-time is
stationary.

To find the general spherically symmetric solution for empty space we
take $dl^2 = \gamma _{\alpha \beta} dx^ \alpha dx^ \beta = e^{2
\lambda}dr^2 + r^2 d{\hat {\bf r}}^2$ where ${\hat {\bf r}}$ is the
unit Cartesian vector $(\sin \theta \cos \phi$, $\sin \theta \sin
\phi, \ \cos \theta)$.  Then $d {\hat {\bf r}}^2 = d \theta^2 + \sin
^2 \theta \, d\phi ^2$ but the advantage of the vector notation is
that no axis for $\theta,\phi$ need be taken.  In spherical
symmetry ${\bf B}_g$ must be radial and divergenceless so Gauss's
theorem gives $|{\bf B}_g|r^2 = Q$ = const which is the field of a
gravomagnetic monopole
$$B_g^r = -Qe^{- \lambda}/r^2. \eqno (3.11)$$
Reinserting ${\bf E}_g = c^2 \nabla\nu$ into (3.8) we have 
$$R_{00} = -\nu'' + \nu'^2 - 2 \nu'/r + \lambda'\nu' + {\textstyle {1
\over 2}}e^{2(\lambda - \nu)} Q^2(cr)^{-4} = 0. \eqno (3.12)$$
To form $P^\alpha \! _\beta$ we need the 3 dimensional Christofel symbols
$$\lambda ^\sigma _{\mu \nu} = {\textstyle {1 \over 2}} \gamma^{\sigma
\eta} (\gamma_{\eta \mu, \nu} + \gamma _{\eta \nu, \mu} - \gamma _{\mu
\nu, \eta}) \eqno (3.13)$$
which are
$${}$$

$\begin{array}{lll} \lambda^\sigma_{\mu \sigma} = {\textstyle {1 \over
2 \gamma}} \, \gamma,_\mu & & \ \ \ \ \lambda^\sigma_{\phi \phi} =
-{\textstyle {1 \over 2}} \gamma^{\sigma \eta} \gamma_{\phi \phi,
\eta} \\ \lambda^\sigma _{rr} \, = \left\{{\ \ \ \ 0 \ \ \ \ \, \ \
\sigma \not= r \atop {\textstyle {1 \over 2}} \gamma^{rr}\gamma_{rr,r}
\ \sigma = r} \right. & & \ \ \ \ \lambda^\sigma_{\theta \theta} \, =
\left\{{0 \atop -{\textstyle {1 \over 2}} \gamma^{rr}\gamma_{\theta
\theta,r}} {\sigma \not= r \atop \sigma = r}\right. \\
\lambda^\sigma_{\phi \sigma} = 0 & & \ \ \ \ \lambda^\sigma_{r \tau}
\, = {\textstyle {1 \over 2}} \gamma^{\sigma \eta} \gamma_{\eta
\tau,r}\end{array}$ $$\eqno (3.14)$$


\noindent(3.6), (3.7) and (3.9) are identically satisfied.  The
surviving equations of (3.10) are
$$R^{rr} = -\nu'' + \nu'^2 + \lambda'\nu - 2\lambda' /r = 0 \eqno (3.15)$$ 
and 
$$\displaylines{R^{\theta \theta} = R^{\phi \phi} = \lambda 'e^{-2
\lambda} - {e^{-2 \lambda} \over r} + {1 \over r} + {1 \over 2}e^{-2
\nu} Q^2c^{-4}r^{-3} + \cr \hfill{} + \nu'e^{-2 \lambda} = 0. \hfill
{(3.16)} \cr}$$ Equations (3.12), (3.15) and (3.16) must be solved for
$\nu$ and $\lambda$.  Eliminating $\nu''$ from (3.12) and (3.15) we
find
$$2(\lambda' - \nu') + {\textstyle {1 \over 2}} e^{2(\lambda - \nu)}
Q^2c^{-4}r^{-3} = 0 \eqno (3.17)$$
which integrates on division by $e^{2(\lambda - \nu)}$ giving
$$e^{-2(\lambda - \nu)} = - q^2r^{-2} + C \eqno (3.18)$$
where $q = Q/2c^2$ which has the dimensions of a length.
Multiplying (3.16) by $e^{2 \nu}$ and using (3.17) and (3.18) we have
$$(C - q^2 r^{-2})(r^{-1} - 2\nu') = q^2r^{-3} + e^{+2 \nu}/r$$
dividing by $e^{2\nu}(C - q ^2 r^{-2})$ we obtain
$$(e^{-2 \nu})' + {1 \over r} \left( {Cr^2 - 2 q^2 \over Cr^2 - q^2}
\right) e^{-2 \nu} - {r \over Cr^2 - q^2} = 0$$
which is linear in $e^{-2 \nu}$ and readily solved by integrating
factor to give
$$e^{-2 \nu} = {1 \over C} - {2q^2 \over Cr^2} + {2 {\overline C}
\over r^2} \sqrt{Cr^2 - q^2} \eqno (3.19)$$
where $C$ and $\overline C$ are both constants.
It follows from (3.18)
$$\gamma_{rr} = e^{2 \lambda} = (C -q^2 r^{-2})^{-1}e^{+2 \nu}. \eqno
(3.20)$$ To get $e^{-2 \nu}$ and $\gamma_{\alpha \beta}$
asymptotically of Schwarzschild form we need $C=1$ and $\overline C =
- \widetilde {m}$, the asymptotic mass $GM/c^2$.  Thus we find
$$g_{00} = e^{-2 \nu} = 1 - 2r^{-2} \left( q^2 + \widetilde
{m} \sqrt{r^2 - \ell ^2} \ \right) \eqno (3.21)$$
$$\gamma_{rr} = (1 - q^2 r^{-2}) e^{+2 \nu} \eqno (3.22)$$
which are the metric components of NUT space.  Notice that when $Q =
2qc^2 = 0$ this reduces to Schwarzschild's metric.  The metric is
completed by taking a vector potential $A_ \alpha$ for the
gravomagnetic field ${\bf B}_g$; any one will do since they are
connected by gauge transformation which merely changes the zero point
of time.  As we saw in Section I it is impossible to choose a
spherically symmetric vector potential but this does not affect the
spherical symmetry of the physics.  A suitable vector potential is
that given in (1.19) which gives us the metric
$$\displaylines{ds^2 = e^{-2 \nu} \left( cdt-2q (1 + \cos \theta) d
\phi \right)^2 - \cr \hfill{}- (1 -q^2/r^2) e^{+2 \nu} dr^2 - r^2 d
{\hat {\bf r}}^2 \hfill{(3.23)} \cr}$$ where $e^{-2 \nu}$ is given by
(3.21).  This metric is more commonly written in terms of the radial
variable $\tilde {r} = \sqrt{r^2 - q ^2}$ because the square roots
disappear leaving an analytic expression, however we have preferred
the variable that makes the surface area of the sphere $4 \pi r^2$ as
in Schwarzschild space.  Of course the metric (3.23) appears to have a
preferred axis but this is illusory because we can switch it into any
direction we like by a gauge transformation, see the discussion under
(1.20).  The horizon where $g_{00}$ changes sign is given by $\tilde
{r} = \widetilde {m} + \sqrt{q^2 + \widetilde {m} ^2}$ at which point
$\gamma_{rr}$ changes sign also as in Schwarzschild space.

NUT-Space was discovered in Ehlers' thesis (1957) and rediscovered by
Newman, Tamburino \& Unti (1963).  It is closely related to Taub's
(1951) metric and their relationship has been beautifully illuminated
by Misner \& Taub (1969).  The fact that NUT space has a gravomagnetic
monopole was found by Demianski \& Newman (1966), who also found a NUT
version of Kerr space.  See also Dowker \& Roche (1967).

\subsection {Orbits and Gravitational Lensing by NUT Space}

The geodesics of NUT space may be determined from $\delta \int ds=0$
using the metric in the form (3.1).  When $ds^2 \not= 0$ we write
$\tau$ for the proper time and when $ds^2 \not= 0$ we replace it by an
affine parameter (which we also call $\tau$).  Varying $\dot t = dt/d
\tau$ and using the fact that the metric is stationary we have 
$$e^{-2
\nu} (c \dot t - A_\alpha \dot x ^\alpha ) = \varepsilon = \ {\rm
constant}. \eqno (3.24)$$
Varying $x^\alpha$ we find
$$\displaylines{\delta x^\alpha \left\{{d \over d\tau} \left[ e^{-2
\nu} (c \dot t - A_\beta \dot x ^\beta) A_\alpha + \gamma_{\alpha
\beta} \dot x ^\beta \right] \right. + \cr \hfill{}
+\left. {\textstyle {1 \over 2}} {\partial \over \partial x^\alpha}
\left[ e^ {-2 \nu} (c \dot t - A_\beta \dot x^ \beta)^2 -
\gamma_{\beta \gamma} \dot x ^\beta \dot x^\gamma \right]
\right\}. \hfill{(3.25)}\cr}$$ Using (3.24) and transferring the
$\varepsilon d A_\alpha/d \tau = \varepsilon \dot x^\beta
\partial_\beta A_\alpha$ term into the second bracket we find the
equation of motion in which ${\bf A}$ only occurs through
$\partial_\alpha A_\beta - \partial _\beta A_\alpha = \eta_{\alpha
\beta \gamma} B^\gamma$ where $\eta_{\alpha \beta \gamma}$ is the
antisymmetric tensor, $\sqrt{\gamma}$ times the alternating symbol.
$$\displaylines{\delta x^\alpha \left[{d \over d\tau} (\gamma_{\alpha
\beta} \dot x ^\beta) + {\textstyle {1 \over 2}} {\partial \over
\partial x^\alpha} (e^ {-2 \nu}) \varepsilon ^2 e^{+4 \nu}\right. -
\cr \hfill - \left. {\textstyle {1 \over 2}} \gamma_{\beta \gamma,
\alpha} \dot x^ \beta \dot x^\gamma - \varepsilon \eta_{\alpha \beta
\gamma} B^ \gamma \dot x^ \beta \Bigr] \right.  = 0 \hfill{(3.26)}
\cr}$$ We now write $\gamma_{\alpha \beta}$ in the form involving the
unit Cartesian vector ${\hat {\bf r}}$,
$$\gamma_{\alpha \beta} dx^\alpha dx^\beta = e^{2 \lambda} dr^2 + r^2
(d{\hat {\bf r}})^2.$$
$\delta {\hat {\bf r}}$ the variation of ${\hat {\bf r}}$ is an
arbitrary small vector perpendicular to ${\hat {\bf r}}$.  Thus making
variations with $r$ fixed we deduce from (3.26) Using (3.11)
for $B_g^r$
$$\delta {\hat {\bf r}} \cdot \left[ {d \over d \tau} \left( r^2 {d
{\hat {\bf r}} \over d \tau} \right) + \varepsilon {d {\hat {\bf r}}
\over d \tau} \times Q {\hat {\bf r}} \right] = 0$$
since $\delta {\hat {\bf r}}$ is an arbitrary vector perpendicular to
${\hat {\bf r}}$ we deduce that
$${\hat {\bf r}} \times {d \over dt} \left( r^2 {d {\hat {\bf r}}
\over d \tau} \right) = {d {\bf L} \over d \tau} = - {d \over d \tau}
\left( \varepsilon Q {\hat {\bf r}} \right)$$
$${\bf L} + \varepsilon Q {\hat {\bf r}} = {\bf j} = \ {\rm const}.
\eqno (3.27)$$
Except for the factor $\varepsilon$ which reduces to $m_0c^2$ in the
non-relativistic case we see that this is precisely the vector
integral (1.14).  Dotting Eq. (3.27) with ${\hat {\bf
r}}$ we find ${\bf j} \cdot {\hat {\bf r}} = \varepsilon Q$ showing
that ${\hat {\bf r}}$ lies on a cone similarly ${\bf L} \cdot {\bf j}
= L^2 = {\bf j}^2 - \varepsilon ^2 Q^2 = \ {\rm const}$ so ${\bf L}$
moves around a cone.  The radial equation of motion is redundant since
we may use the energy and the equation $(ds/d \tau)^2 = U = 1 \ {\rm
or} \ 0$ instead.  $U$ is 1 for time like geodesics and 0 for
light-like ones.

This gives us
$$\varepsilon ^2 e^{+2 \nu} - \dot r ^2 e^{2 \lambda} - L ^2 r^{-2} =
U. \eqno (3.28)$$
To see the geometry of the trajectory we introduce the curvilinear
angle $\varphi$ of $\S$1 measured around the cone's surface.  Then
$r^2 \dot \varphi = L$ so Eq. (3.1) can be integrated by
quadrature
$$\varphi - \varphi_0 = \int {L r^{-2} dr \over \sqrt{\varepsilon ^2
e^{-2 (\lambda - \nu)} - (U + L^2 r^{-2}) e^{-2 \lambda}}}. \eqno
(3.29)$$ In general this integral can not be performed explicitly for
the $\lambda$ and $\nu$ of NUT space even after substitution in terms
of $\tilde {r}$ to make it more analytic.  We therefore turn to the
$r^2 \gg q^2 + \widetilde {m} ^2$ limit well away from the event
horizon.  This is the important case in all gravitational lenses
observed to date.  In that limit the $q^2/r^2$ term in the effective
potential is attractive and therefore of the wrong sign to give the
non-precessing orbits of Section I.  The precession around the cones
is faster than in the classical Kepler + monopole problem by a factor
3/2.  To the first order in $\widetilde {m}/b$ where $b$ is the impact
parameter, we find a bending angle measured like $\varphi$ of $\Delta
\varphi = 4 \widetilde {m}/b$ just as for the Schwarzschild metric;
however the difference is that this angle is measured around a cone
not in a plane.  Again to first order we can find the effect of the
gravomagnetic field by integrating the momentum transfer along the
unperturbed straight line path.  This gives an out-of-plane bending of
$4q/b$; a result that is confirmed by the full NUT space calculation.
Nouri-Zonoz, M. \& Lynden-Bell, D. (1997).  Thus the major effect of
the gravomagnetic monopole $Q$ is to twist the rays that pass it.
While the bending angle is proportional to $b^{-1}$, the effect is
exaggerated when looking down the line toward the NUT lens by the
factor $D_L/b$, so the twist around the lens is $4qD_L/b^2$.  Here
$D_L$ is the distance from the observer to the lens.  The same
exaggeration factor occurs for the normal gravitational bending so for
a source at infinity and an image at $(b, \theta)$ in the plane of the
sky at the lens's distance, the apparent position of the source is
$$(b_s, \theta _s) = \left( b \left( 1 - {4
\widetilde {m} D_L \over b^2} + {8q^2D^2_L \over b^4} \right) , \theta
- {4qD_L \over b^2} \right).$$  
This expression defines a map from
image to source.  From this map one can work out both the shear and
the magnification of a NUT lens in the large impact parameter
r\'egime.  The magnification of area and thus luminosity is
$$db^2/db^2_s = [1-16b^{-4}D^2_L (m^2 + q^2)]^{-1}. \eqno (3.30)$$ A
small circular source will be imaged into an ellipse of axial ratio
$${b^2 +4 \ D_L (m + \sqrt{m^2 + q^2}) \over b^2 + 4 \ D_L (m-
\sqrt{m^2 + q^2})}$$ with the short axis of the ellipse inclined to
the radius at the angle (see Fig. 9)
$$\tan ^{-1} \left( {q \over m+ \sqrt{m^2 + q^2}}
\right).$$  
This is $45^\circ$ for $q \gg m$ and $13^\circ$ for $Q = 2qc^2 =
mc^2$.  This spiral conformation of the images about a NUT lens is
very characteristic.  It is not displayed in normal gravitational
imaging and the gravomagnetic lens due to a rotating object seen pole
on does not show it because the twist of the ray as it approaches such
a lens is cancelled by the opposite twist as it recedes.  Thus the
discovery of a spiral shear field about a lens would indicate the
presence of a gravomagnetic monopole.  Such effects should be looked
for by those studying gravitational lenses.  The expectation must be
small but the reward might be an amazing discovery.

\subsection {Quantization of Gravomagnetic Monopoles and their Classical
Physics}

By analogy with Dirac's argument for the quantization of magnetic
monopoles and charges, Dowker \& Roche (1967), Dowker (1974), Hawking
(1979) and Zee (1985) have suggested quantization of gravomagnetic
monopoles and energy.  Corresponding to Dirac's $Q_me= {\textstyle {1
\over 2}} N \hbar c$ for magnetic $Q_m$, they have $Qm_0 = {\textstyle
{1 \over 2}} N \hbar c$ for gravomagnetic monopole $Q$.  This implies
that both $Q$ and mass $m_0$ are quantized in conjugate units $Q_1$
and $m_1$ obeying $Q_1 m_1 = {\textstyle {1 \over 2}} \hbar c$.
Whereas such ideas are naturally attractive they do not naturally lead
to a self-consistent relativistic theory.  For instance, looking at
the Klein-Gordon equation in NUT space and separating variables with
$\psi \propto e^{i(m \phi + \omega t)}$, one finds an eigenvalue
equation for $\omega$.  The Dirac monopole quantization condition,
$Q(\hbar \omega /c^2) = {\textstyle {1 \over 2}} N \hbar c$ with $N$
an integer, shows us that the only possible eigenvalues $\omega$ are
integer multiples of ${\textstyle {1 \over 2}} c^3/Q$ and the
corresponding energy $\hbar \omega$ is the total energy of the `orbit'
including rest mass.  However this condition conflicts with the
energies of the bound states\footnote {To get definite bound states
one must impose a potential barrier so that the black hole is not
reached.} which are not integer multiples of any unit.  Mueller \&
Perry (1986).  Thus if such ideas are viable at all a more radical
change in basic theory is needed.  In $+++-$ NUT space it does not
appear to be possible to build a consistent quantum theory like
Dirac's magnetic monopole theory. This is what Ross (1983) concluded
and is related to Misner's (1963) finding that NUT space contains closed
timelike lines, with time being periodic every $8 \pi q/c$.  For a
discussion of the energy levels in $++++$ Taub-NUT space, the reader
is referred to the papers by Gibbons \& Manton (1986).  This space was
shown to be relevant to the interactions of monopoles by Atiyah \&
Hitchin (1985).  \S If magnetic monopoles exist, Maxwell's equations
must be changed to include ${\rm div} \ {\bf B} = 4 \pi \rho_ m$,
${\rm Curl} \ {\bf E} + {1 \over c} {\partial {\bf B} \over \partial t} = 4
\pi {\bf j}_m$, where $\rho_m$ is the monopole density and ${\bf j} m$
is the monopole current density.  Such modified Maxwell equations do
not come with a vector potential.  It is natural to ask how general
relativity must be modified to allow for gravomagnetic monopole
densities and currents.  While this is not so obvious we conjecture
the generalization will be to spaces with unsymmetric affine
connections which have non-zero torsion.  It would be interesting to
demonstrate this conjecture as it could introduce a greater degree of
physical understanding of those spaces.

\section {Observability}

Following Kibble's (1980) suggestion that magnetic monopoles would be
a natural consequence of the Big Bang, they have long been sought.  

We have concentrated on the spectra of monopolar atoms and the lensing
properties of gravomagnetic monopoles since these are ways in which,
at least in principle, monopoles might be discovered observationally.
Spectroscopically one may argue that the best place to look is in the
spectra of supernovae, quasars or active galactic nuclei where the
basic Ly $\alpha$ lines of Table I or II might be seen as very weak
absorption lines in very high resolution spectra.  Quasars have the
advantage that these lines will be shifted into the visible.  We have
looked at IUE ultra-violet spectra of Supernova 1987A and seen no
lines at the wavelengths 2774.62 or 2733.78.  More supernovae and
stacked high resolution spectra of quasars should be pursued.
Although in regions of observed magnetic fields the limits obtained
spectroscopically will fall far short of the Parker (1970) bound.  While
the nature of the dark matter that constitutes most mass in the
universe remains unknown, such esoteric possibilities are worth
pursuit.

Searches on Earth have produced one unrepeatable event and a monopolar
observatory under the Grand Sasso that has so far found no monopoles
in Cosmic Rays.  There has been a speculative suggestion, Kephart \&
Weiler (1996), that the leveling up on the numbers of cosmic rays at
the highest energies might be due to monopoles but there is no
confirmation of that idea.  To date the best limit on the numbers of
monopoles in interstellar space comes from the Parker (1970-71) bound.
This arises from the idea that too many magnetic monopoles would
`short out' the galactic magnetic fields that are observed.  A good
general discussion of such limits may be found in the book of Kolb \&
Turner (1991).  For more recent work on monopoles in field theory see
reviews by Olive (1996, 97), and the papers by Sen (1994) and by
Seiberg \& Witten (1994).  More details of the fundamental work on
monopoles in field theory by `t Hooft (1974) and by Polyakov (1974)
can be found in the review by Goddard \& Olive.

\acknowledgments

We thank P. Goddard, D.G. Walmsley and D. Crothers and G.W. Gibbons
for discussions.  M. Mathioudakis brought us the IUE spectrum of
SN1987a and F. McKenna looked for coincidences with spectral lines in
RR Telescopii and pointed out that there is a HeII line at
$\lambda$2733.28A which almost coincides with the predicted
$\lambda$2733.78 line of monopolar hydrogen.  We thank the Director of
the Armagh Observatory where this work began and the Physics
Department of The Queen's University, Belfast, for their hospitality.

\newpage
\begin{figure}
\caption{The circular orbits about a central potential endowed with a
monopole.  The orbits in opposite senses are displaced above and below
the centre of force.  For a Newtonian potential their vertical
separation gradually increases as their radii are increased.  The
orbits with a given angular momentum $|{\bf L}|$ lie on cones with
opening angle $\cos^{-1} (Q_*/|{\bf j}|)$.}
\label{figure1}
\end{figure}
\begin{figure}
\caption{When one of the cones is slit and flattened a gap opens along
the slit.  On the cone itself the sides of this gap are identified.
Orbits which close on a plane will not close on the cone because of
the gap.  As a result they precess.}
\label{figure2}
\end{figure}
\begin{figure}
\caption{An ellipse precessing around a cone of semi angle 70$^{\rm
o}$ making a rosette orbit on it.}
\label{figure3}
\end{figure}
\begin{figure}
\caption{A monopole and its ${\bf B}$ field showing the surfaces $S_1,
S_2$ and $S_3 \equiv S_1-S_2$.}
\label{figure4}
\end{figure}
\begin{figure}
\caption{$j$ values allowed by the conditions $j \geq {\textstyle{1
\over 2}} (|m| + |m - {\textstyle{N \over 2}}|)$.  $j$ cannot be less
than the average of the two faint V lines in the diagram.}
\label{figure5}
\end{figure}
\begin{figure}
\caption{Energy levels for a spinning electron in hydrogen with 0, 1,
2 or 3 Dirac monopoles on its nucleus.  Excepting `isotopic' shifts
due to the changed nuclear mass and relativistic corrections, the
energy levels of the ground states are in the ratio $1:{\textstyle{1
\over 2}}:{\textstyle{1 \over 3}}:{\textstyle{1 \over 4}}$.}
\label{figure6}
\end{figure}
\begin{figure}
\caption{Energy level diagram $E(n,J)$ for $N=1$, hydrogen with one
Dirac monopole on its nucleus.  The nucleus has been assumed to be
fixed.}
\label{figure7}
\end{figure}
\begin{figure}
\caption{Energy level diagram $E(n,J)$ for $N=2$, hydrogen with two
Dirac monopoles on its nucleus.}
\label{figure8}
\end{figure}
\begin{figure}
\caption{Gravitational Lensing by NUT space of a small circular source
at $S$ appears as an inclined ellipse at the image $I$.  Many such
images make a spiral effect around the NUT lens $L$.}
\label{figure9}
\end{figure}

\begin{table}
\caption{Wavelengths in Angstroms of the two Lyman Series, the five
Balmer Series and the eight Paschen Series of hydrogen with $(N=1)$
one Dirac monopole attached to the proton cf. Fig. 7.  The
wavelengths after the dots are those of the series limits.}
\label{table1}
\end{table}
\begin{table}
\caption{Wavelengths in Angstroms of the Lyman, Balmer and Paschen of
hydrogen with two Dirac monopoles attached to the proton cf. Fig. 8
and 6.}
\label{table2}
\end{table}
\end{document}